\documentclass[iop]{emulateapj}

\begin{document}

\title{The Correlation Between H$\alpha$ Emission and Visual Magnitude
During Long-Term Variations in Classical Be Stars}

\shorttitle{H$\alpha$ and V-magnitude Correlations in Be Stars}

\author{T.\ A.\ A.\ Sigut \& P.\ Patel}
\affil{Department of Physics and Astronomy, The University of 
Western Ontario \\ London, Ontario, CANADA N6A 3K7 }
\email{asigut@uwo.ca \\ ppatel54@uwo.ca}

\shortauthors{Sigut \& Patel}
%\shortauthors{Sigut, Patel \& Landstreet}

\slugcomment{Accepted for publication in ApJ January 15, 2013.}

\begin{abstract}

H$\alpha$ equivalent widths and UBV magnitudes are calculated for Be star
disk models that grow in size and/or density with time. We show that these
simple models are consistent with the known Be star classes of positive
and inverse correlations between long-term variations in H$\alpha$
and V magnitude as identified by Harmanec.  We support the conclusion
of Harmanec that the distinction is controlled by the inclination of
the disk to the line of sight. We demonstrate that the strength of
these correlations, particularly those of an inverse correlation where
the system becomes fainter as the H$\alpha$ emission strength grows,
is strongly influenced by the scale height of the inner Be star disk
and the extent of the gravitational darkening of the central B star due
to rapid rotation. This dependence may allow coordinated spectroscopic
and photometric observations to better constrain these poorly known Be
star parameters.

\end{abstract}


\section{Introduction}
\label{sec:intro}

Be stars are defined observationally as non-supergiant B stars that
have shown emission at least once in the Balmer series of hydrogen
(\cite{jas81}; see also reviews by \cite{kh82,slet88,por03}).
Be stars also exhibit an IR excess relative to that expected from
a normal B star \citep{geh74,cote87} and linear continuum polarization
of up to 2\% \citep{mb78,pok79}. All of these observations
are consistent with the presence of an equatorial disk of gas
surrounding the central B star, an idea (``a nebulous ring of
gas") first suggested by \cite{str31}.  This basic picture now has
ample direct interferometric evidence, and the physical model of a
flattened, equatorial disk surrounding the central B star is well
established \citep{dt92,q93,stee95,quir97,tyc05,stee11}. About 17\%
of all non-supergiant B stars in the Milky Way are Be stars, although
the fraction varies widely with spectral type, reaching a maximum of
$\sim\!34$\% at B1 \citep{zor97}. Recently, \cite{tar12} have found that
the fraction of early-type Be stars (B0-B3) peaks in Galaxy clusters
of ages between 12--20~Myr, suggesting that evolutionary status plays
an important role in the Be phenomena.  It also seems well established
that Be stars are more frequent in lower metallicity environments,
such as the Magellanic Clouds \citep{wis06,mar06,mar07b,mar10}.

Be stars exhibit variability on a wide variety of time scales, from
minutes to years/decades \citep{harmanec83,por03}. The origin of short
term variability is usually attributed to pulsations within the B star
photosphere \citep{bad00,hua09}, and there is speculation that pulsations
may play a role in the disk formation \citep{osa99,cra09}. However,
corotating structures, such as photospheric ``clouds," have also
been proposed to explain these variations \citep{harmanec99,bal00}.
Longer term variations seem connected with the circumstellar disk, either
the formation and dissipation of the disk itself, or hydrodynamical
processes occurring within the disk, such as spiral density waves which
lead to the cyclic $V/R$ variations \citep{ok91,hh97,car09}.

One of the striking aspects of Be stars is episodic creation and
destruction of the disk for some Be stars. This is exhibited as a
transition in the observed spectrum from a Be star to a normal B star
and/or to a Be shell star.  Such events present a unique opportunity
to study both the structure of the disk and the physical processes
responsible for its creation.

Even though the process of disk formation is not well understood,
there are observations that seem to show disk formation and
dissipation \citep[eg.\ see][]{wis10}.  Many Be stars show outbursts,
characterized by rapid variability in the emission line profiles. Such
variations are normally interpreted as events of rapid, discrete
mass ejection~\citep{r98a,r98b}, although a definitive observation of
material escaping the stellar surface seems lacking (Harmanec, 2012,
private communication).  These outbursts are thought to be the result of
material that is ejected from the star and are mostly seen in early-type
Be stars.  Stars such as $\omega$~CMa (B2IVe), FV~CMa (B2IV-Vne), FW~CMa
(B2Vne), $\mu$~Cen (B2Vnpe), $\eta$~Cen (B1.5Vne) and 28~Cyg (B2.5Ve) are
examples of early type Be stars with outburst activity \citep{rivinius01}.

Some Be stars are seen to repeat outbursts regularly. In these cases, the
disk is replenished with material every few months to few years, and the
outbursts fill up the inner region of the disk with the material.  If an
outburst is not followed by another one soon, the material falling back on
the star, or propagating outwards in the disk, will create a low density
region near the star. The longer the time before the next outburst, the
emptier the region becomes. This empty region between the star and the
inner edge of the disk makes the disk structure look more like a ring
\citep{meilland06}. This ``inside-out'' mass loss can be seen in many
stars, for example $\eta$~Cen~\citep{rivinius01}, 60~Cyg~\citep{wis10},
and 28 CMa~\citep{ste03}, where observers measure the variability of the
width at the base of the emission lines and/or spectropolarimetric data.

\cite{clark03} give observations for $o$~And (B6IIIpe) which show the
formation and destruction of the disk in an inside-out manner.  In their
15 year data set, the star goes from a normal B star to a Be shell star
and back to normal B star, with the cycle repeating.  This transition
takes around 700 days. Short term outbursts have been noticed which
replenish the disk with more material. \cite{clark03} model one of the
shell episodes, and it is concluded that there is a decrease in density
in the innermost regions of the disk.  This decrease is shown to slowly
propagate outward, indicating an inside-out mass loss in the disk.

Why some B stars, and not others, become Be stars is currently
unclear. Rapid rotation of the central B star seems to play a key role
\citep{por03}, but details are still lacking as a definitive determination
of the actual rotation rates of the central stars is complicated by the
potential effects of gravitational darkening \citep{tow04,cra05,fre05}.
The evolutionary status of the central B star may also play an important
role \citep{eks08,tar12}.

Also unclear is the exact mechanism(s) that creates Be star disks.
Keplerian rotation, now established for Be star disks \citep{hum00,oud08},
suggests viscous decretion as the correct physical process
\citep{lee91,por99}, but how material is feed into the inner edge of the
disk is unknown \citep{cra09,krt11}. It is possible pulsation plays the
critical role \citep{cra09}, and there is tantalizing observational evidence
supporting this view \citep{riv98,hua09}. The role of binarity in the Be
phenomena is also unclear \citep{por03,tar12}, although the majority of
Be stars do not seem to be the result of binary evolution \citep{bad92,vbv97}.

\section{Observed H$\alpha$ and UBV Correlations}
\label{sec:obs}

\cite{harmanec83} describes two classes of long-term variation for
Be stars that can be established from contemporaneous observations of
photometry (such as visual magnitude and UBV colours) and a measure of
the H$\alpha$ emission strength (such as its equivalent width or peak
continuum contrast): positive correlations and negative or ``inverse''
correlations. The distinction between positive and inverse correlations
is thought to be geometric \cite[see][]{harmanec83}, as will be described
shortly.

A positive correlation is seen for most Be stars for which the relevant
observations are available, and it is characterized by a decrease
in the (visual) magnitude of the system with increasing strength of the
Balmer H$\alpha$ emission (such as an increase in its equivalent width
in emission). In a $(U-B)$ vs.\ $(B-V)$ colour-colour diagram, Be stars
with a positive correlation change their luminosity class but not their
spectral type~\citep{harmanec83}.  Increasing H$\alpha$ signifies an
increasing disk, and it therefore seems natural to expect an increase
in the overall brightness of the system, star-plus-disk, as the disk
is built.

Many stars are known to exhibit this positive correlation: $\kappa$~Dra
\citep[B6IIIpe;][]{saad04,juza94}, 60~Cyg \citep[B1Ve;][]{koubsky00},
OT~Gem \citep[B2Ve;][]{bozic99}, V442~And \citep[B2IVe;][]{bozic04},
EW~Lac \citep[B3IVe;][]{floquet00}, $\pi$~Aqr \citep[B1Ve;][]{no77},
QR~Vul \citep[B3Ve;][]{pavlovski83}, $\mu$~Cen \citep[B2Vnpe;][]{dac92},
$\phi$~Per \citep[B2Ve;][]{bozic95}, $\gamma$~Cas
\citep[B0IVe;][]{doazon83}, 28~Tau \citep[B8IVev;][]{hirata00,tanaka07},
and $\omega$~CMa \citep[B2IVe;][]{harmanec98}.

On the other hand, an inverse correlation, as described by
\cite{harmanec83}, is characterized by an an increase in the magnitude of
the system with an increase in the strength of the Balmer emission. In 
the $(U-B)$ vs.\ $(B-V)$ colour-colour diagram, Be stars that show an inverse correlation
move along the main sequence, changing their spectral type, but not
luminosity class~\citep{harmanec83}.

Fewer stars are known to exhibit this inverse correlation: 4~Her
\citep[B9e;][]{koubsky97}, 88~Her \citep[B7pshe;][]{doazan82}, V1294~Aql
\citep[B0Ve;][]{horn82}, and $\eta$~Cen \citep[B1.5IVne;][]{stefl95}.

An inverse correlation is thought to occur when a Be star disk is viewed
more edge-on (i.e.\ at higher inclination angle $i$) \citep{harmanec83}.
The forming disk can then act to reduce the brightness of the system
by blocking the light from the stellar disk, while the small projected
area of the disk on the sky keeps the disk emission to a minimum. If the
critical inclination angle required to observe an inverse correlation
is large enough, inverse correlations will be statistically less
likely to be found than positive correlations.

Hence the basic picture is that for Be stars observed at $i<i_{\rm crit}$,
disk formation should result in a positive correlation, while Be stars
observed at $i\ge i_{\rm crit}$ should exhibit an inverse correlation. Of
course, $i_{\rm crit}$ may depend on spectral type and/or the particulars
of how the disk is being built.  However, $i_{\rm crit}$ must depend to some
extent on the thickness of the disk.  Hence relative numbers of positive
and inverse correlations may be able to constrain, in a general sense, the
thickness of Be star disks. In fact this argument was used originally by
\cite{harmanec83} who noted that the occurrence of positive and inverse
correlations in the long-term variability was, at the time, among the
strongest arguments for flattened Be star disks.  This conclusion is
strengthened by consistency of the correlation classification: if the
record of observations for a given Be star is long enough, the same
type of correlation (positive or inverse) will be repeated \citep{har00}.
As the probability of observing a Be star with inclination $i$ is just
$\sin i\,di$, the probability of observing an inclination $i\ge i_{\rm
crit}$ is $\cos i_{\rm crit}$. Using the number of cited positive and
inverse correlations above, 4 out of 16 Be star systems show an inverse
correlation suggesting $i_{\rm crit}\sim 75^o$. However, this is only
suggestive because the stars used to estimate the
fraction were not selected from a homogeneous sample but simply culled
from the literature.

To date there have been no systematic calculations demonstrating these
correlations, either positive or negative, as a function of viewing
angle $i$. Of particular interest is the case of inverse correlation. At high
inclination angles ($i$ near $90^o$), the disk is viewed edge-on and the
blockage of starlight is caused by the thickness of the disk. While almost
all models of Be star disks assume very thin disks with scale heights
set by hydrostatic equilibrium, there are few direct constraints of this
assumption, and several lines of evidence suggest that the disks may be
much thicker \citep{arias06, zor07b}. In addition, the geometry of the
inverse correlation, namely blockage of the equatorial regions of the
star, suggests that any gravitational darkening of the stellar surface
may also play a critical role. Hence the interplay of disk scale height
and gravitational darkening may allow new constraints to be placed on
these parameters by the detailed study of inverse correlations. It is
the purpose of this paper to provide a preliminary investigation of these
correlations using radiative transfer models for Be star disks.

\section{Calculations}
\label{sec:calc}

The thermal structure of the model Be star disks used in this work
was computed with the {\sc bedisk} code of \cite{sig07}. This code
enforces radiative equilibrium in a photoionized, circumstellar disk
of a prescribed density structure by including heating and cooling
processes for the nine abundant elements (H, He, CNO, Mg, Si, Ca, \& Fe),
each over several ionization stages.  Details of the atomic models and
atomic data, as well as an overview of the {\sc bedisk} code, are given
in \citet{sig07}.  The main energy input into the disk is assumed to be
the photoionizing radiation from the central star.  The older \cite{kur93}
LTE photoionizing fluxes used by \citet{sig07} were replaced with the
newer, non-LTE calculations of \cite{Hubeny07}. To compute the H$\alpha$
equivalent widths and UBV magnitudes of the models, the {\sc beray}
code of \cite{sig11} was used. In all cases, this modelling approach
considers only an isolated Be star and neglects any potential effects
of a binary companion which are outside the scope of the current work.

The density structure of the disk as a function of time was assumed to be of 
the form
\begin{equation} 
\rho(R,Z,t) = \rho_o(t) \left(\frac{R_*}{R}\right)^{n} \,
e^{-\left(\frac{Z}{H}\right)^2} \;,
\label{eq:rho} 
\end{equation} 
where $R$ and $Z$ are the cylindrical co-ordinates for the
axisymmetric disk, and $R_*$ is the radius of the central B star.
The quantities $n$ and $\rho_o(t)$ are adjustable parameters that fix the
density structure of the disk. The disk was assumed to extend from the
stellar photosphere at $R=R_*$ at the inner edge to an
outer disk radius of $R=R_{d}(t)$.  This simple density model has been
very successful in interpreting a wide range of Be star observations
\citep{gie07,tyc08,jon08}.  The disk is assumed to be in Keplerian
rotation and hence rotationally supported in the $R$ direction.

The vertical (or $Z$) dependence of Eq.~(\ref{eq:rho}) contains
the scale height function $H$ which is defined as
\begin{equation}
H=\left(\frac{2R^3\;kT_{HE}}{GM_*\;\mu m_{\rm H}}\right)^{1/2} \;.
\label{eq:scale_height}
\end{equation}
Here $M_*$ is the mass of the central B star, $\mu$ is the
mean-molecular weight of the disk gas, and $T_{HE}$ is an assumed
isothermal disk temperature. The $Z$-dependence of Eq.~(\ref{eq:rho})
and Eq.~(\ref{eq:scale_height}) follow from the assumption that the
disk is in vertical hydrostatic equilibrium set by the $Z$ component of
the stellar gravitational acceleration and an assumed isothermal disk
temperature, $T_{HE}$. In Eq.~(\ref{eq:scale_height}), the mean molecular
weight, $\mu$, was taken to be $0.68$, appropriate for a fully-ionized
hydrogen gas with a 10\% mixture of neutral helium. The isothermal disk
temperature, $T_{HE}$ (which is taken to be a constant for the entire
disk) is further discussed below.

Eq.~(\ref{eq:scale_height}) can be instructively written as
\begin{equation}
\frac{H}{R}=\frac{c_{\rm s}}{V_{\rm K}} \;,
\end{equation}
where $c_{\rm s}$ is the local sound speed in the disk (set by temperature
$T_{HE}$) and $V_{\rm K}$ is the Keplerian orbital velocity at $R$. As
the radiative equilibrium temperatures in the disk give sound speeds
on the order of ten $\rm km\,s^{-1}$ while the orbital speed is many
hundreds of $\rm km\,s^{-1}$, the hydrostatic model predicts disks that
are very thin, often only few percent of the stellar radius near the star.

Note that $T_{HE}$ is only used in Eq.~(\ref{eq:scale_height}) to fix
the vertical density structure of the disk; the actual temperatures
in the disk, $T(R,Z)$, are found by enforcing radiative equilibrium.
\citet{sig09} consider consistent models in which the vertical structure
of the disk is found by integrating the equation of hydrostatic
equilibrium in a manner consistent with the radiative equilibrium disk
temperatures: this treatment eliminates the need for the parameter
$T_{HE}$. In general, \citet{sig09} find that vertical disk scale
heights are generally overestimated by Eq.~(\ref{eq:scale_height}),
using the typical value of $T_{HE}\approx\,0.6\,T_{\rm eff}$, in the
inner regions of the disk due to a cool zone that tends to form in the
equatorial plane of the disk near the star for disks of sufficient density
(i.e.\ $\rho_0$). Nevertheless, as will be shown in a later section,
disks using Eq.~(\ref{eq:scale_height}) with $T_{HE}=0.6\,T_{\rm eff}$
are already sufficiently thin to have significant consequences for the strength of the predicted
H$\alpha$-V magnitude correlations.\footnote{\cite{msj12} demonstrate
that the effect of gravitational darkening on
the disk temperatures can result in scale heights even smaller than those found
by \cite{sig09}. Hence the use of Eq.~(\ref{eq:scale_height}) in the
current work should be considered an upper limit to the thicknesses of
disks in hydrostatic equilibrium.}

The thermal disk models are thus described by the spectral type of the
central B star (which is assumed to fix the stellar mass, radius and
effective temperature) and the parameters in Eq.~(\ref{eq:rho}) that fix
the density of the disk: $\rho_o$, the base density (in $\rm g\,cm^{-3}$),
$n$, the power-law index, and $R_{\rm d}$, the outer disk radius.
The isothermal disk temperature used in Eq.~(\ref{eq:scale_height}),
$T_{\rm HE}$,  is also a parameter of the models, and variations of this
parameter are the subject of Section~\ref{subsec:enhanced}.

In the present work, two spectral types were considered for the central
B stars: B1V and B5V, representing an early and later-type Be star. The
assumed fundamental parameters are given in Table~\ref{tab:bstars}.
Be stars are, of course, known as rapid rotators and hence the role of
gravitational darkening of the stellar photosphere must be considered.
This effect will be explicitly considered in Section~\ref{subsec:grav}.

\begin{deluxetable}{lrr}
\tablewidth{0pt}
\tablecaption{Parameters adopted for the central B stars.\label{tab:bstars}}
\tablehead{
\colhead{Parameter} & \colhead{B1V} & \colhead{B5V}}
\startdata
Mass ($M_{\sun}$)       & 13.2  &  5.9 \\
Radius ($R_{\sun}$)     & 6.4   &  3.9 \\
Luminosity ($L_{\sun}$) & $1.4\cdot 10^4$ & $6.9\cdot 10^2$ \\
$\rm T_{eff}$(K)        & $25\,000$ & $15\,000$ \\
$\log(g)$	        & $4.0$     & $4.0$ \\
$\rm T_{HE}$    & $15\,000$ & $9\,000$ \\
\enddata
\tablecomments{The mass and radius calibrations are taken from \cite{cox00}.}
\end{deluxetable}

As this paper is concerned with the optical signature of disk formation
and dissipation, some prescription for these processes is required.
One-dimensional, hydrodynamic simulations that evolve the disk surface
density as a function of time within the framework of viscous decretion
have reached a high degree of sophistication \citep{car10,hau12} and
have been impressively used to model specific Be stars in considerable
detail (see, for example, \cite{car09} for $\zeta$~Tau). \cite{car12}
have recently used the observed decline of the V-magnitude of 28~CMa
during a disk dissipation phase to infer a value of $\alpha=1.0\pm0.2$
for the disk viscosity parameter using detailed hydrodynamic and radiative
transfer modelling.  However this current paper is focused mainly on the
geometrical origin of the correlation classes noted in the previous
section and the influence of the disk scale height and gravitational
darkening of the central stars on these correlations. In addition,
the models here are not meant to represent any specific Be star in
detail. For these reasons, a very simple model was used to build the
disk as a function of time (variations of this basic description are
discussed at the end of this section). The overall density of the disk
was assumed to linearly increase over a time $\beta$ to a maximum value
of $\rho_{\rm max}$, namely \begin{equation} \label{eq:db1} \rho_o(t) =
\rho_{\rm max} \, \left( \frac{t}{\beta} \right) \;.  \end{equation} The
outer edge of the disk was assumed to expand linearly with time with
a fixed radial expansion speed $v_r$ as \begin{equation} \label{eq:db2}
R_{\rm d}(t) = v_r \, t \;.  \end{equation} For both spectral types, we took
$\rho_{\rm max} = 10^{-10}\,\rm g\,cm^{-3}$ and $\beta=365\,$days. For
the B1V model, we took $v_r = 5\,\rm km\,s^{-1}$ and for the B5V model, we
took $v_r = 3\,\rm km\,s^{-1}$. These choices ensure that the disk sizes
expressed in terms of the stellar radius are the same for both models.
Hydrodynamical simulations suggest outflow velocities in viscous disks
of a few $\rm km\,s^{-1}$ \citep{lee91,por99}.  In addition, in the
density law of Eq.~(\ref{eq:rho}), $n$ was taken to be 3.5 and $T_{HE}$
was set to $0.6\,T_{\rm eff}$.  The growth of the disk with time for
both models is given in Table~\ref{tab:disk_rho}.  After one year, the
disk around the B1V model reaches 1.05~AU in radius and has a mass of
$3.25\cdot10^{-9}\,\rm M_{\sun}$; the B5V model reaches 0.63~AU with a
mass of $6.50\cdot10^{-10}\,\rm M_{\sun}$.  We note that both of these
models result in quite substantial Be star disks, with low-inclination
H$\alpha$ equivalent widths approaching 40\,\AA\ in emission.

\begin{figure}
\plotone{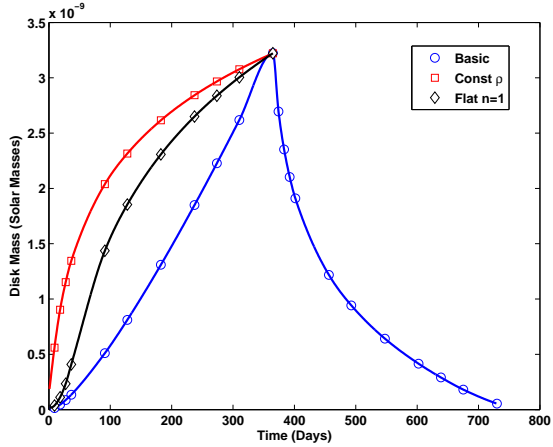}
\caption{Disk mass as a function of time for the B1V model. The maximum
disk mass is $3.25\cdot10^{-9}\,\rm M_{\sun}$ at one year. The circles
represent the basic model described by Eqs.~(\ref{eq:db1}) and
(\ref{eq:db2}). The squares represent the constant $\rho_0$ model and
the diamonds, the flat $n=1$ model, as described in the text.  The disk
mass as a function of time for the B5V model is very similar except that
the maximum disk mass is $6.50\cdot10^{-10}\,\rm M_{\sun}$ after one year.
\label{fig:disk_mass_b1v}} \end{figure}

To dissipate the disk as a function of time, a inner hole was evacuated
in the final density model of the build phase at $t=365\,$ days starting
at the inner edge of the disk at $1\,R_*$.  The outer edge of the hole
was assumed to propagate outward at speed $v_r$ so that the disk sizes of
Table~\ref{tab:disk_rho} also represent the outer radii of the evacuated
hole at time $t+365\;$ days.  Inside this hole, the density was reduced
by a factor of $10^3$.  Thus the mass of the disk is reduced by a factor
of $10^3$ over the duration of the dissipation phase (one year) and the
final density model for both disks was $n=3.5$ and $\rho_0=10^{-13}\,\rm
g\,cm^{-3}$.  This density is so low that the H$\alpha$ line of the
model is indistinguishable from the photospheric (absorption) line
and the IR excess is $<\,1$\%.  Thus the assumption is that the disk
is lost ``inside-out," a direction consistent with many observations
\citep{rivinius01,clark03,meilland06}.  The disk mass as a function of
time for the B1V model is shown in Figure~\ref{fig:disk_mass_b1v}. Note
that in the dissipation stage, the sharp density contrast at the outer
edge of the evaluated hole was smoothed somewhat to improve the radiative
transfer solution there.

%The growth is essentially linear in the build phase, but non-linear
%in the dissipation phase. This latter behaviour is due to an smoothing
%of the inner edge of the disk (or outer edge of the evacuated cavity)
%to improve the radiative transfer treatment in the disk.

\begin{deluxetable}{rrr}
\tablewidth{0pt}
\tablecaption{Disk density and radius as a function of time for the B1V and
B5V models. \label{tab:disk_rho}}
\tablehead{
\colhead{Time} & \colhead{$\rho_0$} & \colhead{$R_{\rm d}$} \\
(days) & $({\rm g\,cm^{-3}})$ & ($R_*$) }
\startdata
  9.1  & $2.5\,10^{-12}$ & $ 1.9$ \\
 18.2  & $5.0\,10^{-12}$ & $ 2.8$ \\
 27.4  & $7.5\,10^{-12}$ & $ 3.6$ \\
 36.5  & $1.0\,10^{-11}$ & $ 4.5$ \\
 91.3  & $2.5\,10^{-11}$ & $ 9.8$ \\
127.7  & $3.5\,10^{-11}$ & $13.4$ \\
182.5  & $5.0\,10^{-11}$ & $18.6$ \\
237.3  & $6.5\,10^{-11}$ & $23.9$ \\
273.7  & $7.5\,10^{-11}$ & $27.5$ \\
310.3  & $8.5\,10^{-11}$ & $31.0$ \\
365.0  & $1.0\,10^{-10}$ & $36.3$ \\
\enddata
\tablecomments{
%$R_*=6.42\,R_{\sun}$ for the B1V model and $3.9\,R_{\sun}$ for the B5V model. 
The final disk masses after one year are $3.25\cdot10^{-9}\,\rm M_{\sun}$ (B1V)
and $6.50\cdot10^{-10}\,\rm M_{\sun}$ (B5V).}
\end{deluxetable}

The calculation of the H$\alpha$ line profiles was done with the {\sc
beray} code \citep{sig11}. Here the radiative transfer equation is
solved along a series of rays through the star-plus-disk system. For
rays terminating on the stellar surface, photospheric LTE H$\alpha$
profiles were computed with the code of \cite{bar03}. To obtain the
V-magnitude of the system, a visual SED was computed for each model and
the V-transmission curve of \cite{johnson66} was used.

Only a single {\sc bedisk} model is required for the disk-building
sequence as the disk can truncated at the appropriate outer radius in the
{\sc beray} code. However, an individual {\sc bedisk} model is required
for each time in the disk loss sequence. As the disk is evacuated
inside-out, the growing size of the inner gap changes
the photoionizing radiation field reaching the remaining parts of the
disk. Hence a consistent thermal structure for the disk is required in
each case, and the temperature structure of the disk changes along the
loss sequence. This is further discussed in Section~\ref{subsec:enhanced}.

In addition to the basic set of disk models described above by
Eqs.~(\ref{eq:db1}) and (\ref{eq:db2}), two alternate assumptions about
the growth of the disk mass during the build phase were considered. In the
first variation, Eq.~(\ref{eq:db1}) was replaced with the assumption of
a constant base density, $\rho_o(t)=\rho_0$, for all times, where $\rho_0$
was taken to be $10^{-10}\,\rm g\,cm^{-3}$. In the second variation,
the constant density $\rho_o(t)=\rho_0$ assumption was used along with the
requirement that in the inner region of the disk where $R\le 5\,R_*$,
the power-law index $n$ in Eq.~(\ref{eq:rho}) was $n=1.0$. An index of
$n=3.5$ was retained for $R> 5\,R_*$. In this case, to give a model with
a final disk mass consistent with the other two models, $\rho_0$ was set
to $2.7\cdot\,10^{-12}\,\rm g\,cm^{-3}$.  Figure~\ref{fig:disk_mass_b1v}
compares the disk mass as a function of time in the build phase for these
two variations with the basic model previously described. Both of these
variations result in much more massive disks at early times but result in
the same total disk mass at the end of the disk build phase at 365~days.

\section{Results}
\label{sec:results}

\subsection{Basic Model}
\label{subsec:basic}

To establish a reference set of models, the disks surrounding the
stars were assumed to be in strict vertical hydrostatic equilibrium
with no gravitational darkening of the central star. For typical
stellar parameters, Eq.~({\ref{eq:scale_height}}) predicts very thin
disks with $(H/R)\sim 0.05$ in the inner regions. In addition, it is
well known that Be stars are rapid rotators and hence gravitational
darkening is potentially an important effect.  This will be discussed
in Section~\ref{subsec:grav}.

As discussed in the Section~\ref{sec:obs}, stars showing a positive
correlation show a decrease in visual
magnitude with an increase in the equivalent width (EW) of H$\alpha$.
Stars showing an inverse correlation show an
increase in visual magnitude with an increase in the EW of H$\alpha$. For
the calculations in this report, instead of the visual magnitude itself,
the change in visual magnitude ($\Delta V$) is used which is defined
as the difference between the visual magnitude of a given model and the
visual magnitude value for the last model of the disk loss phase at 730
days. As noted in the previous section, this last model shows no evidence
of disk emission and can be taken to represent the isolated B star.

\begin{figure}
\plotone{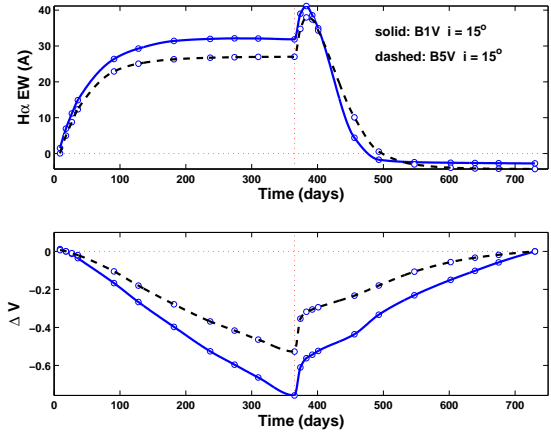}
\caption{H$\alpha$ equivalent width (\AA, top panel) and change
in V-magnitude (bottom panel) as a function of time for the B1V
(solid lines) and B5V (dashed lines) models seen nearly pole-on at
$i=15^{o}$. All models assumed a thin, hydrostatic disk with $H$ set by
Eq.~(\ref{eq:scale_height}) and gravitational darkening of the central B
star was not included. The disk grows for one year and then dissipates over
one year as described in the text. 
\label{fig:panel_i15}} \end{figure}

Consider first the B1V and B5V models seen nearly face-on at
$i=15^o$. Figure~\ref{fig:panel_i15} shows the disk growth ($1-365$
days) and dissipation ($366-730$ days) phases of both models.
Both spectral types behave in a very similar manner. The top panel
of the figure shows the change in EW of H$\alpha$ emission line as
a function of time,{\footnote{We adopt the non-standard convention
that a positive EW denotes emission and a negative EW, absorption.}}
while the bottom panel shows the change in visual magnitude ($\Delta
V$). In the top panel, the H$\alpha$ equivalent width increases with
time as the density and the outer radius of the disk increase. It can
be seen that the strength of H$\alpha$ reaches its maximum, $\approx
30\,$\AA, by essentially $\sim\,100$ days, when the outer radius of
disk has reached $\sim\,10\;R_*$ (Table~\ref{tab:disk_rho}). This
is consistent with the observed sizes of Be star disks resolved with
optical interferometry using H$\alpha$~\citep{tyc05,grun06}. When the
disk dissipation stage starts at 365 days, there is a temporary increase
in the H$\alpha$ equivalent width for another $\sim\,15$ days before
entering a steep decline. This rather counter-intuitive result occurs
because the equivalent width is a measure of the strength of H$\alpha$
relative to the local continuum, i.e.\ \begin{equation} {\rm EW} \equiv
\int \left(\frac{F_{\nu}-F_c}{F_c}\right)\; d\nu \,, \end{equation}
where $F_{\nu}$ is the flux at frequency $\nu$ and $F_c$ is the flux
of the reference continuum. Because of the very large optical depths
in H$\alpha$, it is formed over a wide region of the disk extending
out to $\sim\,10\;R_*$. The optical continuum, on the other hand,
forms closer to the star. So in an inside-out dissipation scenario,
the strength of the reference continuum is reduced more quickly
than that in the line, and this leads to an increase in the equivalent
width, even though $F_{\nu}$ declines during this phase. This `bump'
is over by $\sim\,400$ days, by which time the inner evacuated zone has
expanded to $\sim\,3.5\;R_*$.

The growth and dissipation phases do not mirror one another because
of the inside-out nature of the dissipation. In the growth phase, the
base density ($\rho_0$) and outer radius ($R_{d}$) of the disk both grow
linearly in time. However, during the dissipation phase, only the extent
of the inner evacuated region grows linearly with time, leading to an
inner cavity in the disk. By 730 days, the disk has $\rho_o=10^{-13}
\rm g\,cm^{-3}$ and extends to $36\,R_*$. This disk is so rarefied
that there is essentially no emission due to the disk in H$\alpha$, and
its EW represents the (negative) absorption width of the photospheric
line. Note that the first calculated model in the growth phase is after
9 days at which point the disk has grown to $1.9\,R_*$. This is the
reason the initial equivalent width in the Figure~\ref{fig:panel_i15}
is slightly larger (i.e.\ affected by emission) than the last model in
the dissipation sequence.

The bottom panel of Figure~\ref{fig:panel_i15} shows the change in the
V-magnitude over the growth and dissipation
phases. The reference model for the colour differences is chosen to be the
last model in the dissipation sequence (730 days) because, as noted above,
it exhibits a pure photospheric spectrum which is uncontaminated by the
disk. The sense of the visual magnitude changes are always brightening
in the growth phase and dimming
in the dissipation phase (with one brief exception noted below). 
As the system is viewed nearly face-on, there
is little obscuration of the star by the disk, and hence the disk always
adds to the brightness of the system. The very rapid decline in the V
magnitude of the system at the onset of the dissipation phase reflects
the above discussion about the formation of the optical continuum close
to the star, within a few stellar radii.

In the correlation classification scheme of Section~\ref{sec:obs}, both
models (B1V and B5V) would be an instances of a positive correlation as
the system brightens as the disk (and the H$\alpha$ emission strength)
grows. Note, however, there would be a brief window of an inverse
correlation during the initial phase of the inside-out dissipation, as
the system would grow fainter while H$\alpha$ continued to increase
during the `bump' between 365 and 380 days.

\begin{figure}
\plotone{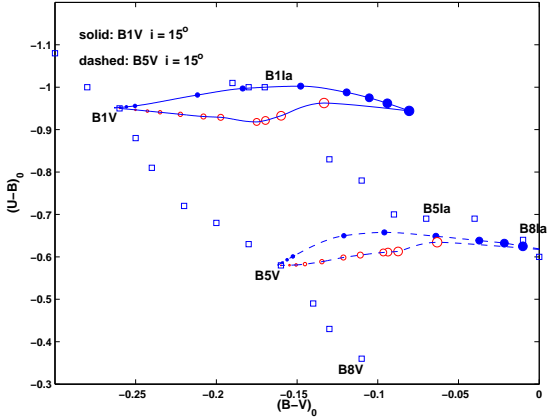}
\caption{The movement of the B1V (solid lines) and B5V (dashed lines)
models in an optical colour-colour diagram as the disk grows (filled blue
circles) and dissipates (open red circles). The models assumed a thin,
hydrostatic disk and gravitational darkening was not included. The
models are viewed nearly pole-on at $i=15^{o}$. The circles grow and
diminish in size following $\rho_o(t)$. The squares are the observed
normal (star only) colours for main sequence (V) and supergiant (Ia)
stars from \cite{fitzgerald70}.  \label{fig:colcol_i15}} \end{figure}

Another interesting way to look at this sequence of models is in a
colour-colour diagram of $(U-B)$ versus $(B-V)$ where $UBV$ are
the familiar Johnson photometric colours. The evolution of the entire sequence 
for the B1V and B5V models is shown in Figure~\ref{fig:colcol_i15}. 
To calibrate the colours,
the colours of the last model in the dissipation phase (730 days)
were forced to be the colours of a normal B1V or B5V star as tabulated
by~\cite{fitzgerald70}. The additive constants were then applied to
every model in the sequence. Note that as the first model is after
9 days of disk growth, it will not lie exactly on the main sequence
reference point. For both modes, the growth track is from the
main sequence, nearly horizontal in the figure, towards the supergiant
sequence, resulting in models that have nearly constant $(U-B)$
colours but progressively redder $(B-V)$ colours.  The tracks are such
that the systems move from the V region, horizontally, towards the Ia
region of the colour-colour diagram; this is the observed behaviour for
positively correlated systems in which movement in the colour-colour
diagram is often categorized as the system ``changing its luminosity
class between V and Ia without any substantial change of its spectral
type" \citep{harmanec83}.  This behaviour is a reflection of the nature
of the additional continuum emission from the disk. Typically, the
continuum excess for Be stars grows with longer wavelengths (because it
is principally due to free-free emission). Hence the $U$ colour is hardly
affected, while the $V$ colour is most affected (as noted above). As the
star is essentially unobscured for $i=15^o$, this simple fact gives the
horizontal motion in the colour-colour diagram.

\begin{figure}
\plotone{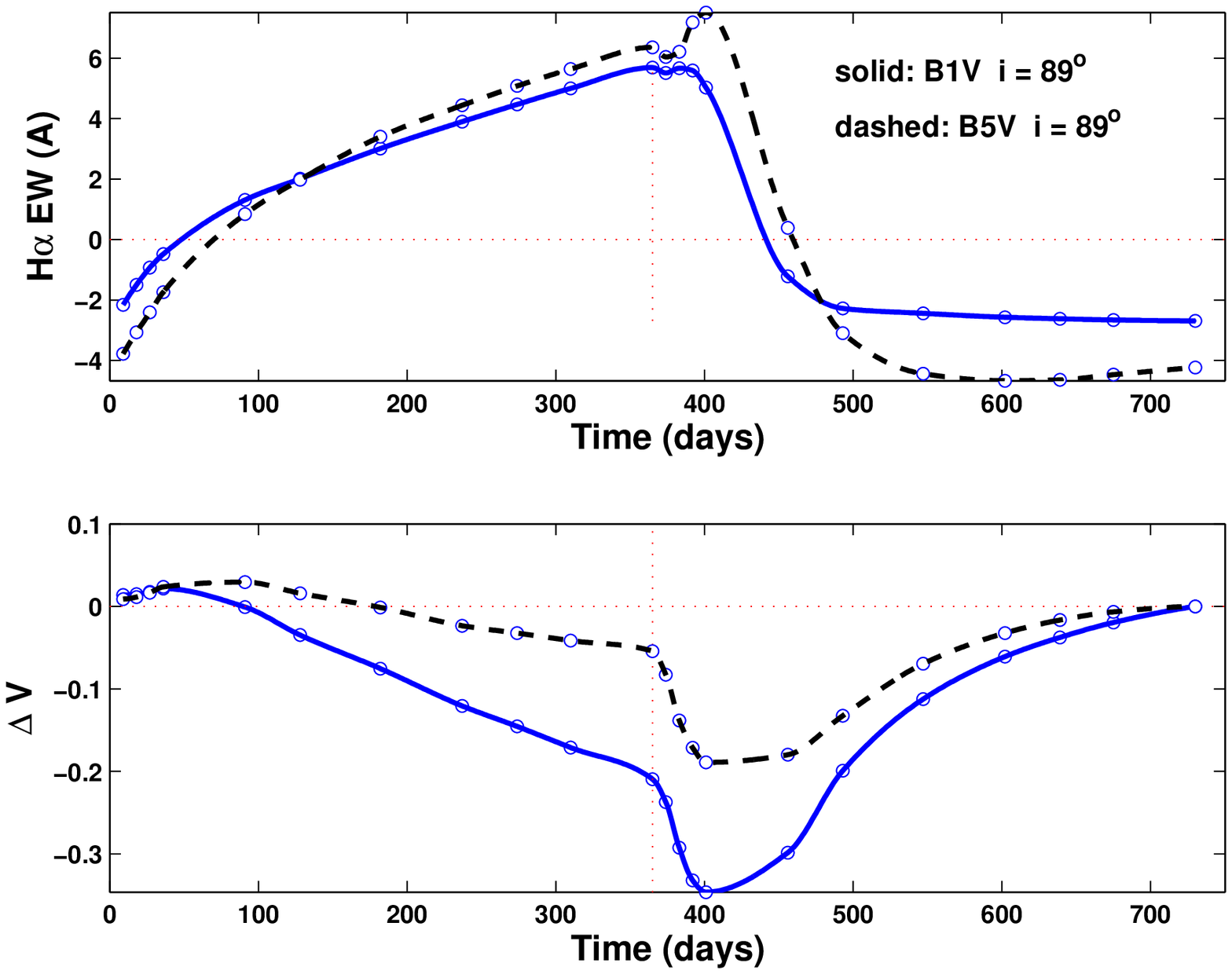}
\caption{Same as Figure~\ref{fig:panel_i15} but for edge-on disks seen
at $i=89^{o}$.  \label{fig:panel_i89}} \end{figure}

We now turn to the same set of models, B1V and B5V, but for a nearly
edge-on disk seen at $i=89^o$. In this case, the disk can obscure the
star's photospheric light, and this leads to new effects shown
in Figure~\ref{fig:panel_i89}. The top panel shows the change in the
H$\alpha$ EW, and it again grows with time as the disk is built. It starts
at negative values this time (i.e.\ net absorption) due to the much weaker
disk emission at $i=89$ degrees and the effect of shell absorption. The
photospheric equivalent widths are represented by the equivalent widths at
730 days.  In this series, the H$\alpha$ equivalent width grows steadily
with time and reaches a maximum in emission of $\approx\,6\,$\AA\ at 365
days. In the edge-on case, the maximum H$\alpha$ EW is much smaller than
in the face-on case, as expected from the much smaller projected area of
the optically thick disk on the sky. The bottom panel shows the effect on
the V magnitude. For times less than about 100 days, the system is fainter
than the star itself, reflecting the obscuration by the disk. However,
the effect is very small (a few hundredths of a magnitude), and this is
a direct consequence of the very thin nature of a purely hydrostatic disk.

In the colour-colour diagram corresponding to the $i=89^o$ models
(Figure~\ref{fig:colcol_i15}), more inclined tracks are seen. In this
case, the U colour is affected by absorption by the disk, making it larger
and hence $(U-B)$ smaller. Observationally, inversely correlated systems
``move along the main sequence changing its spectral but not luminosity class"
\citep{harmanec83}. This behaviour is clearly reproduced for the B5V model
at $i=89^o$ during its disk build phase. However, the B1V model does not
move along the main sequence, and this may be related to the very weak inverse
correlations predicted by this set of basic models with hydrostatic disks.

\begin{figure}
\plotone{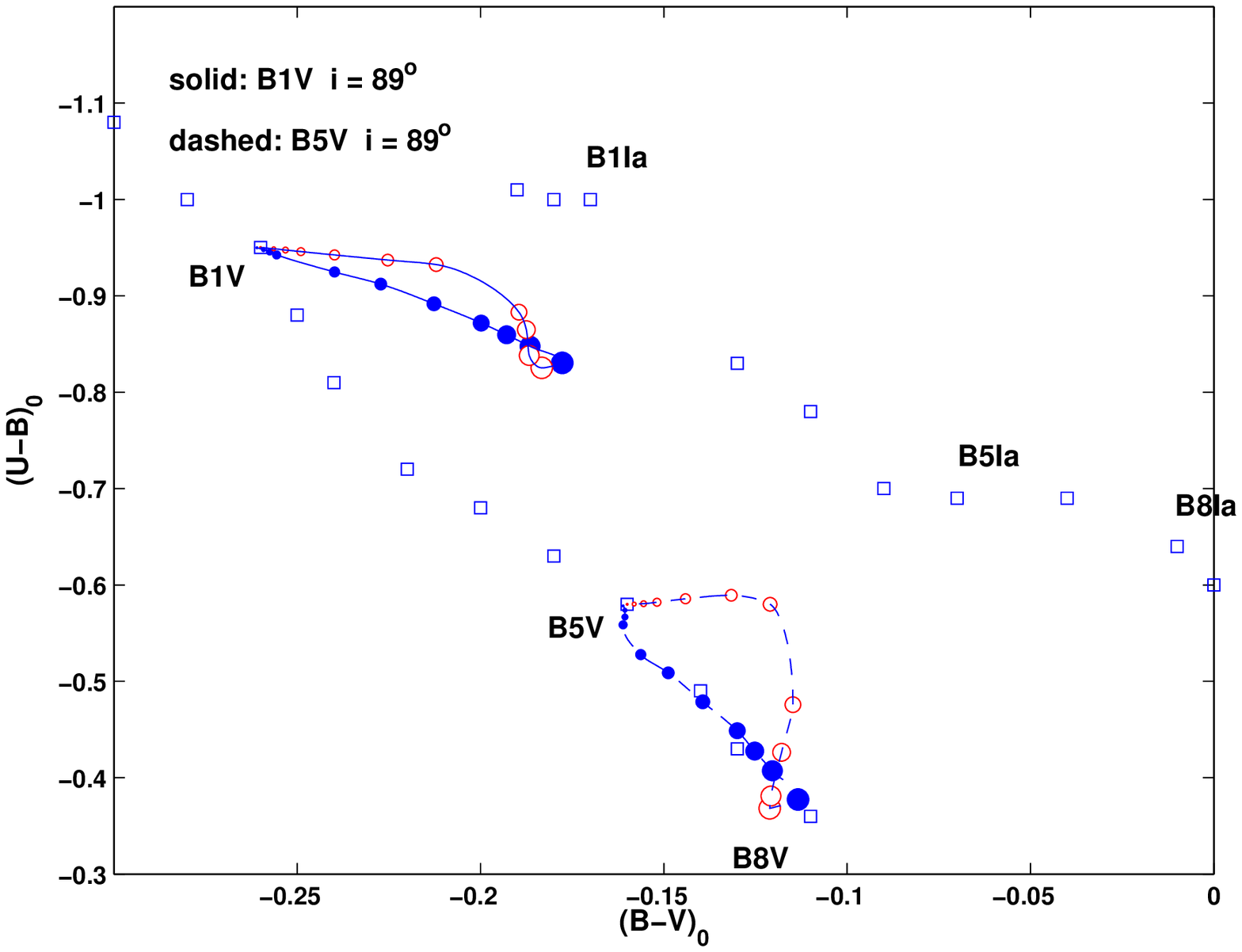}
\caption{Same as Figure~\ref{fig:panel_i15} but for edge-on disks seen
at $i=89^{o}$.  \label{fig:colcol_i89}} \end{figure}

\subsection{Enhanced Disk Scale Heights}
\label{subsec:enhanced}

In the previous section, we presented the optical signature of a disk
build and dissipation sequence. The case of positive correlation
is well represented by the $i=15^o$ models of both spectral types.
The overall $V$ magnitude of the system increases by approximately half
a magnitude over the build phase of the disk. In addition, the movement
in the colour-colour diagram is as observed: towards the giant sequence
at nearly constant spectral type.

However, the case of an inverse correlation is less well represented.
In the edge-on model, for which an inverse correlation is expected,
only a very small magnitude increase is seen during the initial build
phase of the disk, amounting to a few hundredths of a magnitude. Observed
inversely correlated systems typically show magnitude changes of a tenth
of a magnitude or more. Behaviour in the colour-colour diagram is more
as observed; there is a tendency to evolve along the main sequence,
changing the system's spectral type.

One feature of the reference models leading to the very small magnitude
increase during edge-on observation is the very thin nature of the disks.
Almost all calculations of Be star disks, either hydrodynamical or
thermal/radiative, assume a geometrically thin disk with the scale
height set by vertical hydrostatic equilibrium. As previously noted,
this scale height is the ratio of the local sound speed to the local
orbital speed.  As Be star disks are relatively cool and the orbital
speeds are Keplerian and large, the disk scale heights are predicted
to the very small; near the star, the scale height is often only a
few percent of the stellar radius.  Some have questioned whether this
assumption is really correct, and there is some circumstantial evidence
to support enhanced inner disk scale heights \citep{arias06,zor07b}.

\begin{figure}
\plotone{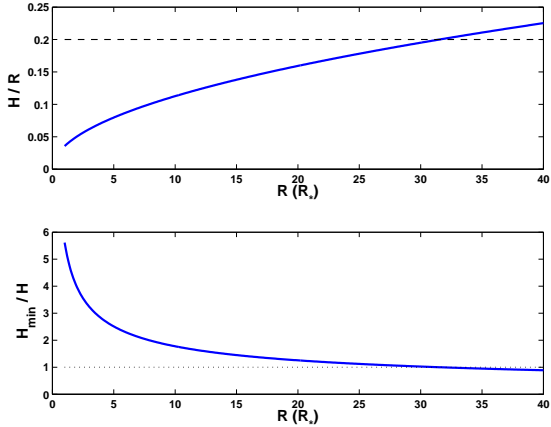}
\caption{Top panel: hydrostatic scale height (Eq.~\ref{eq:scale_height})
for the B1V model assuming $T_{HE}=15\,000\;$K. The scale height $(H/R)$ at 1 $R_*$
is predicted to be $\sim0.04$. $(H/R)_{\rm min}=0.2$, adopted for some models in this work,
is shown as the dashed line. Bottom panel: the disk scale height
enhancement over the hydrostatic value obtained adopting a minimum value 
for $(H/R)$ of $0.2$. The dotted line is only a guide for the eye.
\label{fig:H_b1v}} \end{figure}

To examine the sensitivity of the predicted magnitude changes to
the inner disk scale height, we have adopted a simple model in which
the disk scale height is not allowed to fall below a minimum value,
$(H/R)_{\rm min}$, which we take to be $0.2$. The effect on the disk
scale height of the B1V model is shown in Figure~\ref{fig:H_b1v}.  As can
be seen, this assumption results in a disk of nearly constant scale
height $(H/R)=0.2$ until $\approx\,30$ stellar radii. Near the star,
the implied scale height enhancement is a factor of $\approx\,5$. Given
this modification, the temperature structure of the disk, the $H\alpha$
EWs, and the optical $UBV$ magnitudes and colours were recomputed.

\begin{figure}
\plotone{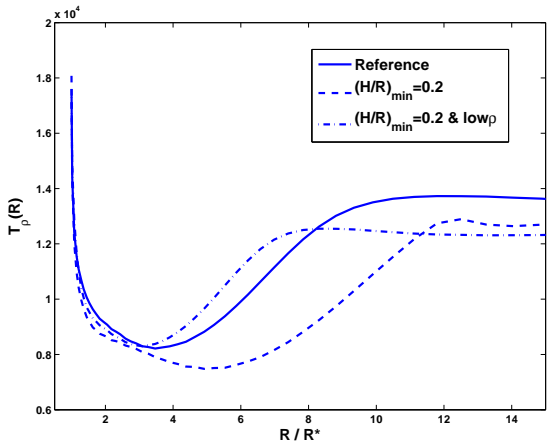}
\caption{Comparison of the equatorial disk temperature between the
reference B1V model (solid line) and a model with $(H/R)_{\rm min}$
set to $0.2$ (dashed line).  Also shown is the temperature structure
of a $(H/R)_{\rm min}=0.2$ model with $\rho_0$ reduced so that the
total disk mass equals that of the reference B1V model (dash-dot
line). \label{fig:tcomp}} \end{figure}

One subtlety of the enhanced scale height models is that simply increasing
the scale height over the reference model will result in disks that are
more massive. To avoid this effect, we have reduced the $\rho_0$ of the
enhanced models at each time in Table~(\ref{tab:disk_rho}) by the amount
required to match the disk mass of the reference model.  Hence the
disk mass of the enhanced scale height models as a function of
time match those given in Figure~\ref{fig:disk_mass_b1v}. We note that
the total emission measures of the disks now agree much better as well
(within 30\%). Another important effect of this reduction in $\rho_0$
is that the temperature structure of the enhanced scale height models
are now much closer to those of the basic reference models.  This is
shown in Figure~\ref{fig:tcomp}. Without the reduction in $\rho_0$, the
enhanced scale height model for B1V has a much more extensive inner cool
zone in the equatorial plane as compared to the reference hydrostatic
model. The enhanced scale height model with reduced $\rho_0$ shows much
better agreement.

In this initial set of enhanced scale height calculations, we ignored
gravitational darkening; this is potentially an important effect and
it will be modelled in Section~\ref{subsec:grav}. Hence the results here
can be used for Be stars rotating at less than $\approx\,70$\% of their
critical velocity. Instead of presenting the entire sequence of models, as
before for the reference models, we will focus solely on the correlation
between the optical magnitude and H$\alpha$ equivalent width during the
disk build sequence from 1 to 365 days.

\begin{figure}
\plotone{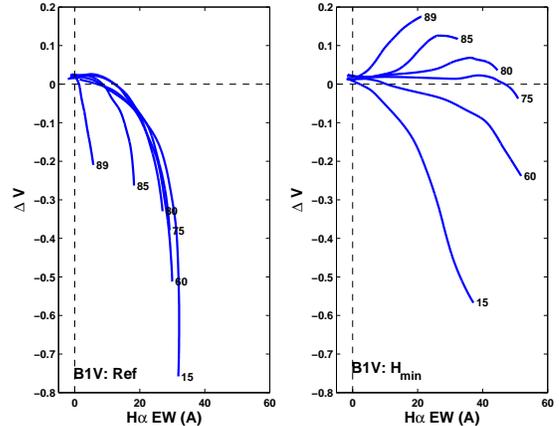}
\caption{Relation between the change in V-magnitude and the H$\alpha$
equivalent during the disk building phase (1-365 days) for the B1V model.
Each curve is labelled by the inclination of the system to the line of
sight. The left panel shows the case of pure hydrostatic disks and the
right panel shows the case of models with enhanced vertical scale heights.
\label{fig:b1v_0p00}} \end{figure}

\begin{figure}
\plotone{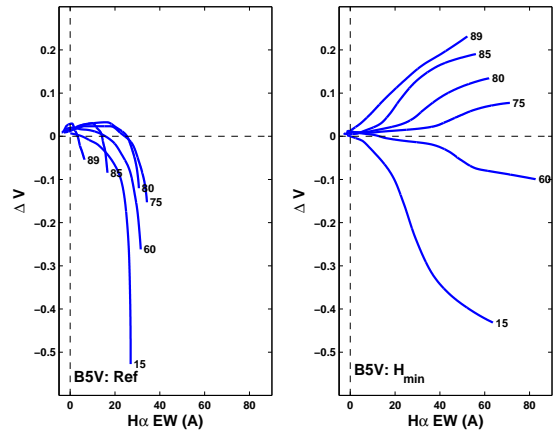}
\caption{Same as Figure~\ref{fig:b1v_0p00} except for the B5V model.
\label{fig:b5v_0p00}} \end{figure}

The results are shown in Figure~\ref{fig:b1v_0p00} for the B1V models
and Figure~\ref{fig:b5v_0p00} for the B5V models.  The contrast with
the reference hydrostatic models is striking. The thin, hydrostatic models
show only a very small decrease in the brightness of the system while the
enhanced scale height models clearly show strong inverse correlations
for inclinations above about $\approx\,75^o$. Perhaps the most
significant difference is that increases in the visual magnitude of the
system occur even for the largest values of H$\alpha$ EW. In the
pure hydrostatic models, all of the curves turn over at intermediate EW,
and the system becomes brighter. From the enhanced scale height models,
the dividing viewing inclination between positive and negative or inverse
correlation is seen to be $\approx\,75^o$. Hence the choice of $(H/R)_{\rm min}=0.2$
seems to give an $i_{\rm crit}$ in agreement with the rough estimate of
Section~\ref{sec:obs} based on the observed numbers of positive and negative
correlations. Nevertheless, this should not be over interpreted; a more
homogeneous sample of correlations coupled with detailed modelling of
individual stars is required for a firm conclusion on enhanced disk scale
heights for Be stars.

\subsection{Alternate Disk Growth Models}
\label{sec:alt}

One feature of the disk growth model implied by Eqs.~(\ref{eq:db1})
and (\ref{eq:db2}) is that the disk density during the early
growth phases, $t<100\,$days, is very low, and hence there is not a
large impact on the system's spectral energy distribution. To see the
effect of a much denser disk during the early growth stages, we return
to the two alternate disk growth assumptions outlined at the end of
Section~\ref{sec:calc}. Note that both of these alternate models assume
a thin disk with the vertical scale height set by hydrostatic equilibrium
via Eq.~(\ref{eq:scale_height}); these are not enhanced scale height
models.

\begin{figure}
\plotone{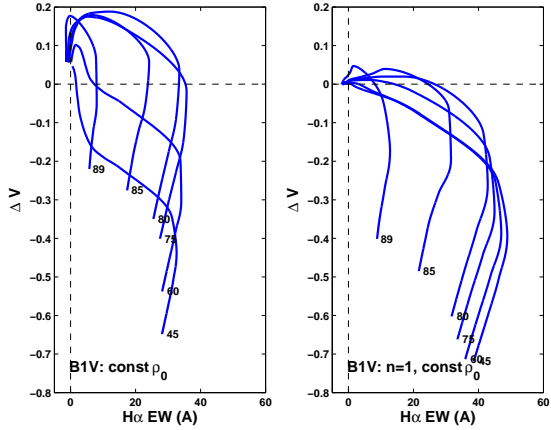} \caption{Same as
Figure~\ref{fig:b1v_0p00} for the B1V model but implementing
the alternate disk growth scenarios of Section~\ref{sec:alt}. The left panel is
the constant $\rho_0$ disk, while the right panel has constant $\rho_0$ and $n=1$ in the inner
region. \label{fig:b1v_alt}} \end{figure}

\begin{figure}
\plotone{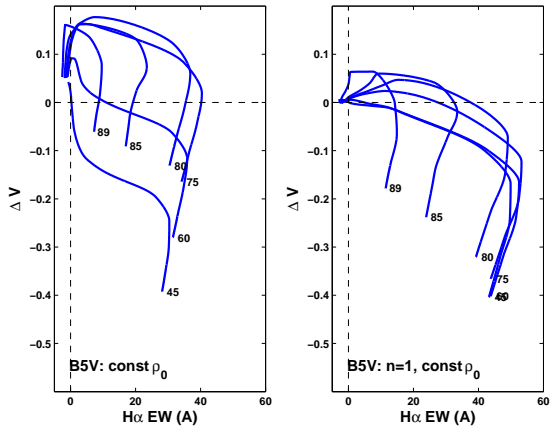}
\caption{Same as Figure~\ref{fig:b1v_alt} but for the B5V model.
\label{fig:b5v_alt}} \end{figure}

In the first alternate model with the base disk density set to a constant
$\rho_o(t)=10^{-10}\,\rm g\,cm^{-3}$, the outer radius of the disk simply
grows with time. For this model, we have recalculated the change in
V-magnitude and H$\alpha$ equivalent width during the growth phase
($t\le 365\,$ days), and the results are shown in the left-hand panel
of Figure~\ref{fig:b1v_alt} for the same viewing angles as the previous
figures. As can be seen from the figure, the $\Delta\,V$ at early times is
much larger because of the denser disk. The period of inverse correlation,
however, is confined to the very initial times when the H$\alpha$ EW
is still small. For later times, and larger values of the H$\alpha$
EW, the correlation flattens out and then turns over for all viewing
angles. The case of the B5V model is shown in the left-hand panel of
Figure~\ref{fig:b5v_alt} and is very similar to the B1V model, except
that changes in the V magnitude are somewhat smaller.

In the second alternate disk growth model, a constant base disk
density was again used, and in addition, a much shallower density
drop-off in the equatorial plane for $R_d\le 5\,R_*$
was assumed by setting $n=1$ in Eq.~\ref{eq:rho} over this range. 
To ensure that the fully built
disk had the same mass as the previous models, the base disk density was taken
to be a constant $\rho_o(t)=2.7\,\cdot\,10^{-12}\,\rm g\,cm^{-3}$; the disk mass as
a function of time is also shown in Figure~\ref{fig:disk_mass_b1v}. The
results in the $\Delta\,V$-H$\alpha$ diagram are shown in the right-hand
panel of Figure~\ref{fig:b1v_alt} for the B1V model. In this case,
weak inverse correlations are predicted for inclinations $i\ge\,80^o$,
but the dimming of the system is small, $\Delta\,V\approx\,0.05\,$mag. In
addition, like the previous case of $\rho_o(t)=10^{-10}\,\rm g\,cm^{-3}$,
the inverse correlation is confined to small H$\alpha$ EW; the relation
flattens out and turns over into a positive correlation for larger
EWs. A very similar result is obtained for the B5V model, as shown in
the right-hand panel of Figure~\ref{fig:b5v_alt}.

\subsection{Gravitational Darkening}
\label{subsec:grav}

Be stars are known to be rapid rotators \citep{por96,yud01}, as first
suggested by \cite{str31}, although how close they are as a population
to critical rotation is still uncertain \citep{tow04,cra05,mei12}. Rapid
rotation leads to gravitational darkening of the stellar surface in
which the temperature varies with latitude, and the shape of the star is
no longer spherical. According to von Zepiel's theorem \citep{vonz24},
the local effective temperature should be proportional to the effective
gravitational acceleration at each latitude, and this effect causes the
equatorial regions of a rapidly rotating star to be much cooler than the
poles. Gravitational darkening, both as the distortion of the stellar
surface and as a variation of temperature over the stellar surface, now
has ample, direct interferometric verification \citep{vanb12}. These
studies have lead to the conclusion that while gravity darkening
is present, von Zepiel's theorem as applied to stellar atmospheres
by \cite{coll66} overestimates the temperature variation. Recently,
\cite{elr11} have derived a new version of gravitational darkening that
seems in better agreement with observations; we adopt their formalism.

Gravitational darkening is potentially an important effect in the current
work, particularly for the case of an inverse correlation because the
blocked equatorial region is now much cooler and hence fainter. One
might expect that the size of the inverse correlation will depend on
the extent of gravitational darkening.

We have used the formalism of \cite{elr11} to compute the variation of
temperature with latitude over the stellar surface. Given the exploratory
nature of the present calculations, however, we have neglected the
distortion of the stellar surface. A useful term for this approximation is
spherical gravitational darkening (SGD). We have also neglected the effect
of gravitational darkening on the thermal structure of the circumstellar
disk. This latter effect is extensively discussed by \cite{msj11} and
even at critical rotation, the effect would be
at most a 15\% reduction in the average disk temperature.

\begin{figure}
\plotone{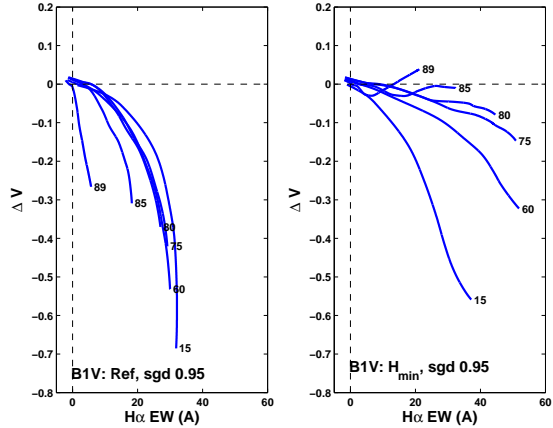} \caption{Same as
Figure~\ref{fig:b1v_0p00} for the B1V model but implementing
spherical gravitational darkening assuming the star is rotating
at 95\% of its critical velocity.  \label{fig:b1v_0p95}} \end{figure}

\begin{figure}
\plotone{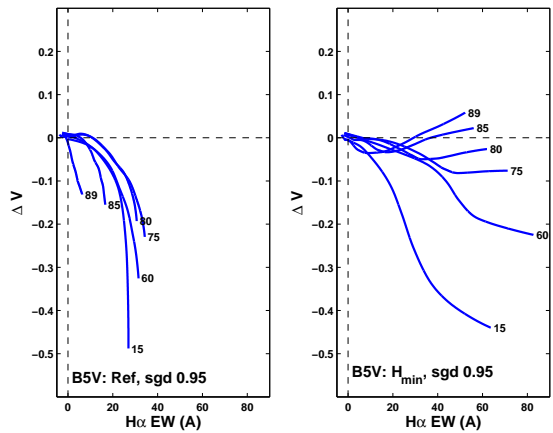}
\caption{Same as Figure~\ref{fig:b1v_0p95} but for the B5V model.
\label{fig:b5v_0p95}} \end{figure}

Figure~\ref{fig:b1v_0p95} shows the effect of SGD, with the central star
rotating at 95\% of its critical velocity, on the B1V model, both with a
thin, hydrostatic disk (left panel) and the thicker $(H/R)_{\rm min}=0.2$
disk (right panel). In the case of the thin, hydrostatic disk, almost all
trace of the inverse correlation is gone. In the $(H/R)_{\rm min}=0.2$ case,
the inverse correlation is now seen only for higher inclinations ($i\ge
80^0$) and only for latter times in the disk-building process. A very
similar result is seen in Figure~\ref{fig:b5v_0p95} for the B5V model.

As a result, the existence of an inverse correlation depends strongly
on the extent of gravitational darkening and hence, on the rotation
rate of the central B star of the system. Note that these results would
only contradict the existence of inverse correlations if all Be stars
were rapid rotators. While the exact rotational distribution of Be stars
remains unclear, it seems unlikely that all of them are critical rotators,
at least for the earlier spectral types \citep{cra05}.

\section{Discussion}
\label{sec:discuss}

In this work, simple parametrized models for Be star disk growth
and dissipation are investigated to see if they can reproduce the
known classes of positive and inverse correlations discovered by
\cite{harmanec83}. We find that these simple models can demonstrate
both positive and inverse correlations between the system's change in
visual magnitude and the magnitude of the H$\alpha$ emission. We also
demonstrate the scale height of inner circumstellar disk and the extent
of gravitational darkening of the central star's surface play key roles
in controlling the magnitudes of the predicted correlations, particularly
the inverse ones.

What do these calculations demonstrate? It is important to keep in mind
that only parametrized, ``toy" models of disk growth and dissipation
were used, and hence it is difficult to draw general conclusions. However
the calculations performed suggest that inverse correlations are
more easily predicted for Be star disks with enhanced scale heights
around B stars that are not significantly gravitationally darkened.
Because the predicted correlations depend so strongly on these two
aspects, gravitational darkening and disk scale height, it would be
highly desirable to analyze a well determine set of contemporaneous
magnitude and H$\alpha$ measurements through either a Be star growth or
dissipation phase to determine to what extend the disk scale height and/or
extent of gravitational darkening can be constrained. In the analysis
of real observations, a much more realistic hydrodynamical description
of disk growth would be preferable and such models are now available
\citep[e.g.\ ][]{car12,hau12}.

\acknowledgments
This work is supported by the Canadian Natural Sciences and Engineering
Research Council (NESRC) through a Discovery Grant to TAAS. PP would
like to thank J.\ D.\ Landstreet for additional financial support through
his NSERC Discovery Grant. We thank J.\ D.\ Landstreet and Peter Harmanec
for very helpful comments on the manuscript.

%
%%%%%%%%%%%%%%%%%%%%%%%%%%%%%%%%%%%%%%%%%%%%%%%%%%%%%%%%%%%%%%%%%%%%%%%%%
%

\end{document}